\documentclass[12pt,preprint]{aastex}
\input psfig.sty
\shortauthors{F. Frontera et~al.}
\shorttitle{GRB011211 with BeppoSAX}

\def\sax{{\em BeppoSAX\/}}   

\def\etal{{\it et al. }}

\def\ergcms{\mbox{ erg cm$^{-2}$ s$^{-1}$}}
\begin{document}

\title{A decreasing column density during the prompt emission from
GRB000528 observed with \sax}

\author{F.~Frontera\altaffilmark{1,2},
L.~Amati\altaffilmark{2},
D.~Lazzati\altaffilmark{3},
E.~Montanari\altaffilmark{1},
M.~Orlandini\altaffilmark{2},
R.~Perna\altaffilmark{4},
E.~Costa\altaffilmark{5},
M.~Feroci\altaffilmark{5},
C.~Guidorzi\altaffilmark{1},
E.~Kuulkers\altaffilmark{6}
N.~Masetti\altaffilmark{2},
L.~Nicastro\altaffilmark{7},
E.~Palazzi\altaffilmark{2},
E.~Pian\altaffilmark{8}, 
L.~Piro\altaffilmark{5}
}

\altaffiltext{1}{Physics Department, University of Ferrara, Via Paradiso
 12, 44100 Ferrara, Italy; frontera@fe.infn.it}

\altaffiltext{2}{Istituto Astrofisica Spaziale e Fisica Cosmica, section of 
Bologna, CNR, Via Gobetti 101, 40129 Bologna, Italy}
%
%\altaffiltext{3}{Space Research Organization in the Netherlands,
% Sorbonnelaan 2, 3584 CA Utrecht, The Netherlands}
%
\altaffiltext{3}{Institute of Astronomy, University of Cambridge, 
Madingley Road, Cambridge CB3 0HA, UK}

\altaffiltext{4}{Department of Astrophysics, University of Princeton 
University, Peyton Hall, Ivy Lane, Princeton, NJ 08544-0001, USA}

\altaffiltext{5}{Istituto Astrofisica Spaziale e Fisica Cosmica, CNR, Via Fosso del 
Cavaliere, 00133 Roma, Italy}
\altaffiltext{6}{ESA, ESTEC, Keplerlaan 1, NL-2200 AG Noordwijk, Netherlands}

\altaffiltext{7}{Istituto Astrofisica Spaziale e Fisica Cosmica, section of Palermo, 
CNR, Via U. La Malfa 153, 90146 Palermo, Italy}

\altaffiltext{8}{Osservatorio Astronomico di Trieste, INAF, Trieste, Italy}

%
%AUTHOR LIST AND ORDER ARE PROVISIONAL

\begin{abstract}
We report observation results of the prompt X-- and $\gamma$--ray
emission from GRB~000528. This event was detected with the Gamma-Ray
Burst Monitor and one of the two Wide Field Cameras aboard the \sax\
satellite.  The GRB was promptly followed on with the \sax\ Narrow
Field Instruments and with ground optical and radio telescopes.  The X--ray
afterglow source was identified, but no optical or radio counterpart
found. We report here results
from the prompt and afterglow emission analysis. The main feature derived
from spectral evolution of the prompt emission is a high hydrogen-equivalent 
column density with evidence of its decrease with time. We model this 
behavior in terms of a time-dependent photoionization
of the local circum-burst medium, finding that a compact and dense
environment is required by the data. We also find a fading 
of the late part of the 2--10 keV prompt emission which is consistent 
with afterglow emission. We discuss this result in the light
of the external shock model scenario.

\end{abstract}

\keywords{gamma rays: bursts --- gamma rays: observations --- X--rays:
general ---absorption lines}

\section{Introduction}
\label{s:intro}

The prompt emission of long duration ($>$2 s) Gamma Ray Bursts (GRBs)
in the broad 2--700 keV energy band, which was accessible to the \sax\
wide field instruments (Wide Field Cameras, Jager et al. 1997
\nocite{Jager97}, and Gamma Ray Burst Monitor, Frontera et al. 
1997\nocite{Frontera97}), has
been demonstrated to provide information not only on the X--ray
emission mechanisms of GRBs but also on their environments (e.g.,
Frontera et al.  2000\nocite{Frontera00}, Frontera
2004a\nocite{Frontera04a}).  The properties of these environments are
of key importance to unveil the nature of the GRB progenitors, which
is a still open issue.  Many observational results point to the
collapse of massive stars, in particular to the explosion of type Ic
supernovae, as the most likely progenitors of long GRBs. However the
simultaneity of the GRB event and supernova explosion (hypernova
model, e.g., Paczynski 1998\nocite{Paczynski98}), in spite of the
convincing results obtained from GRB0303289 \cite{Stanek03,Hijorth03}, 
cannot be considered a solved problem for all GRBs.  A two-step process, 
in which first the massive star
gives rise to a supernova explosion with the formation of a neutron
star, and thus, after some time, to a GRB from the delayed collapse of
the NS to a BH (supranova model, Vietri \& Stella
1998\nocite{Vietri98}) or from a phase transition to quark matter
\cite{Berezhiani03} cannot be robustly excluded. At least in some
cases (e.g., GRB990705, Amati et al. 2000 \nocite{Amati00}; 
GRB011211, Frontera et al. 2004b\nocite{Frontera04b}, GRB991216, Piro et
al. 2000\nocite{Piro00}), in which X--ray transient absorption
features or emission lines have been observed and found consistent
with an Iron-enriched environment typical of a previous supernova
explosion, the two--step process is favored (Lazzati et al. 2001a,
2001b, 2002\nocite{laz2001a,laz2001b,laz2002}).

The properties of the GRB environment can also be derived from the
study of the continuum spectrum of the prompt X--ray emission, which
is expected to show a low-energy cutoff in case of a high column density along
the line of sight. If the absorbing material is located between a
fraction of a parsec to several parsecs from the GRB site, the cutoff is expected to
fade as the GRB evolves. This is due to the progressive
photoionization of the gas by the GRB photons (e.g., B\"ottcher et al.
1999\nocite{Boettcher99}; Lazzati \& Perna
2002\nocite{LazPer02}). This is an effect analogous to the one
originally proposed by Perna \& Loeb (1998)\nocite{perlo98} in the
optical domain with resonant lines. Actually, such cutoffs, with also
evidence of their decreasing behaviour with time from the GRB onset,
have already been detected (GRB980506, Connors \&
Hueter 1998\nocite{Connors98}; GRB980329, Frontera et
al. 2000\nocite{Frontera00}). Lazzati \& Perna (2002)\nocite{LazPer02}
interpreted them as evidence for presence of overdense regions in
molecular clouds with properties similar to those of star formation
globules.

Now we report the discovery of another case of absorption.
It has been found in the X--ray spectrum of the prompt emission of
GRB000528, in the context of a systematic investigation
\cite{Frontera04c} performed to study the spectral evolution of all
GRBs jointly detected with WFCs and GRBM. GRB000528, detected with the
GRBM \cite{Guidorzi00} and localised with the \sax\ WFC
\cite{Gandolfi00}, was promptly followed-up with the \sax\ Narrow
Field Instruments (NFI), with the discovery in the WFC error box of a likely
X--ray afterglow candidate of the burst \cite{Kuulkers00}.  However
neither an optical or radio counterpart was found (lowest upper limit
to the $R$ magnitude of 23.3, Palazzi et al.  2000\nocite{Palazzi00};
3.5$\sigma$ upper limit of 0.14 mJy at 8.46 GHz, Berger \& Frail
2000\nocite{Berger00}).

In this paper we report our findings and their interpretation.  We
also discuss the follow--up X--ray observations performed with the
\sax\ NFIs and their implications.

\section{\sax\ observations} 
\label{s:obs}

GRB000528 was detected with the \sax\ WFC No. 2 and GRBM
\cite{Gandolfi00} on 2000 May 28 at 08:46:21 UT. The GRB position was
determined with an error radius of $2'$ (99\% confidence level) and
was centered at $\alpha_{2000} = 10^{\rm h}45^{\rm m}06.3^{\rm s}$,
and $\delta_{2000} = -33^\circ58'59''$ \cite{Zand00}.  The burst was
then followed--up with the NFIs two times, the first time (TOO1) from
May 28.71 UT to May 29.57 UT (from 8.25 to 28.90 hrs after the main event),
and the second time (TOO2) from May 31.65 UT to June 1.49 UT (from 3.284 to
4.124 days from the main event).  Two previously unknown sources, the
first one (S1) designated 1SAX~J1045.5$-$3358, with coordinates
$\alpha_{2000} = 10^{\rm h}45^{\rm m}28^{\rm s}$, and $\delta_{2000} =
-33^\circ 58'14''$, the second one (S2) designated
1SAX~J1045.1$-$3358, with coordinates $\alpha_{2000} = 10^{\rm
h}45^{\rm m}08^{\rm s}$, and $\delta_{2000} = -33^\circ 59'26''$ were
detected with MECS and both found to have disappeared in TOO2
\cite{Kuulkers00}. However only the S2 position (uncertainty radius of
1.5$'$) was found consistent with the WFC error box (see also Hurley et al.
2000\nocite{Hurley00}).
  
\begin{figure*}[!t]
\epsscale{1.1}
\plotone{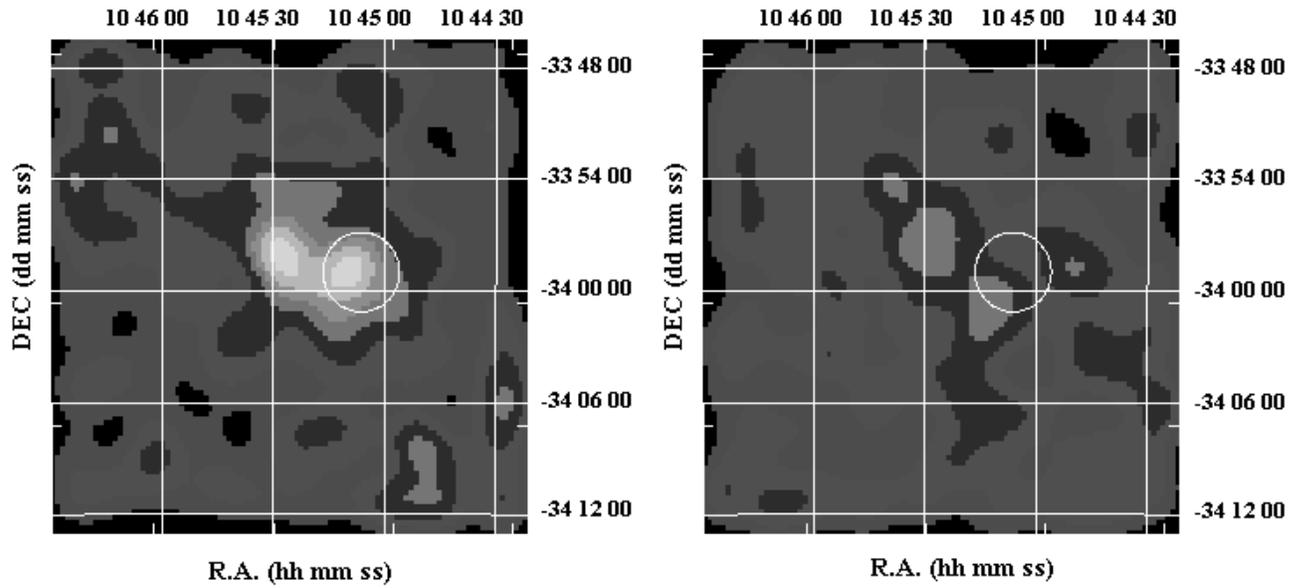}
\vspace{1.5cm}
\caption{2--10 keV image of the sky field obtained with MECS in the two
\sax\ follow--up observations of GRB000528. The first TOO started 8.25 hrs after
the burst, the second 3.284 days later.  The error box obtained with the
WFC 2 is also shown.}
\label{f:image}
\end{figure*}

We have reanalysed the MECS and LECS data. The exposure times of the
two instruments to the GRB direction are 8084 s (LECS) and 26589~s
(MECS) in TOO1, and 6905~s (LECS) and 31210~s (MECS) in TOO2.  The
images of the sky field derived with MECS 2+3 during the two TOOs are
shown in Fig.~\ref{f:image}, with superposed the WFC error box. The
two sources reported by Kuulkers et al.\nocite{Kuulkers00} are clearly visible
in the left image. The probability that the excesses found are due to
background fluctuations is $\sim 7 \times 10^{-7}$.  
The source S1, which is out of the WFC error box, is coincident with a radio source, which 
was already present in the NRAO VLA Sky Survey (NVSS) at the time of the 
GRB occurrence. It was again detected by Berger \& Frail (2000)\nocite{Berger00} 
during the search of the radio counterpart of GRB000528. This source is 
also detected in the second observation (chance probability of $\sim 5 
\times 10^{-4}$), with a flux variation by a factor 1.7 
from the first to the second TOO. However the source S2 
is detected only during the first TOO, it is consistent with the WFC position, and 
undergows a flux decrease larger than a factor 3 from the first to the 
second TOO (see Section~\ref{s:aft}). As already concluded by Kuulkers et al. 
(2000)\nocite{Kuulkers00}, we assume this object as the likely afterglow source
of GRB000528.

\section{Data analysis and results}
The spectral analysis of both the prompt and afterglow emission was
performed with the {\sc xspec} software package \cite{Arnaud96}, version 11.2. All
the reported parameter uncertainties are given at 90\% confidence level.

\subsection{Prompt emission}
\label{s:prompt}

Data available from GRBM include 1~s ratemeters in the two energy
channels (40--700~keV and $>$100~keV), 128~s count spectra
(40--700~keV, 225 channels) and high time resolution data (up to
0.5~ms) in the 40--700~keV energy band.  WFCs were operated in 
normal mode with 31 channels
in 2--28~keV and 0.5~ms time resolution.  The burst direction was
offset by 8.0$^\circ$ with respect to the WFC axis. With this offset,
the effective area exposed to the GRB was $\approx$~340~cm$^2$ in
the 40-700~keV band and 70~cm$^2$ in the 2--28~keV energy band.
The background in the GRBM energy band was estimated by linear
interpolation using the 250~s count rate data before and after the
burst.  The WFC spectra were extracted through the Iterative Removal
Of Sources procedure (IROS
\footnote{WFC software version 105.108}, e.g.  Jager et al. 1997
\nocite{Jager97}) which implicitly subtracts the contribution of the
background and of other point sources in the field of view.

\begin{figure}[!t]
\epsscale{0.8}
\plotone{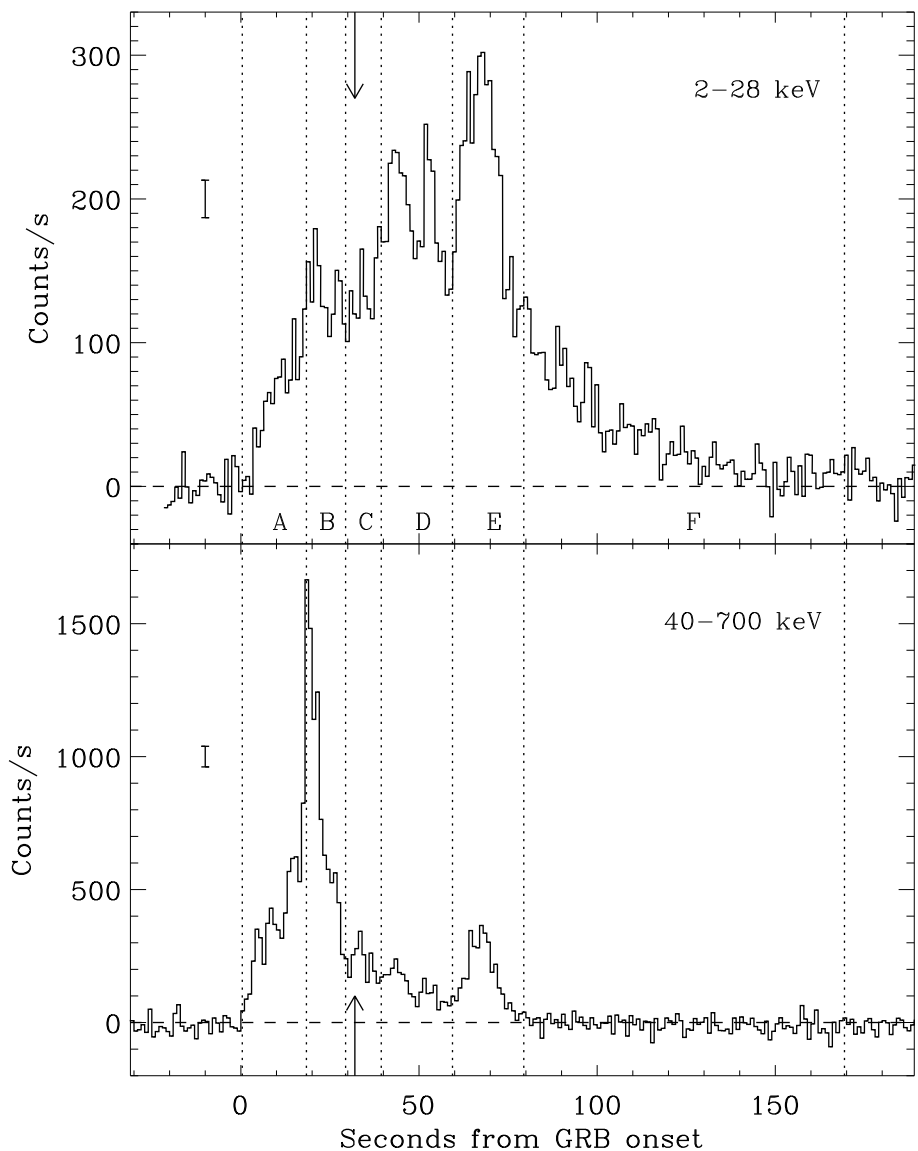}
\vspace{1.5cm}
\caption{Light curve of GRB000528 in two energy bands, 2--28 keV (WFC),
and 40--700 keV (GRBM), after background subtraction. The zero abscissa corresponds 
to 2000 May 28, 08:46:21 UT. The time slices over which the 2--700 keV spectral 
analysis was performed using the 1~s GRBM counters are indicated by vertical 
dashed lines. The two time slices (named 1 and 2 in the text) for which 
also the 225 channel GRBM spectra were available are separated by 
a vertical arrow.}
\label{f:lc}
\end{figure}

Figure~\ref{f:lc} shows the light curve of GRB000528 in two energy
bands. In X--rays (top panel of Fig.~\ref{f:lc}) it shows a seemingly
long rise with superposed various small pulses until it achieves the
highest peak at a level of 300 cts~s$^{-1}$ after $\sim 70$~s from the
onset. Then the X--ray flux starts to smoothly decay and ends after
$\sim 120$~s from the flux peak.  In gamma--rays the light curve
(bottom panel of Fig.~\ref{f:lc}) is totally different. In
correspondence of the first of the secondary pulses of the X--ray light
curve we find the most prominent gamma-ray pulse, where the GRB achieves the
maximum count rate in the 40--700 keV band of 1700
counts~s$^{-1}$. Instead, in correspondence of the peak flux in
X--rays we only find a secondary pulse of the gamma--ray light
curve. Smaller and shorter duration pulses are also visible in the
gamma--ray light curve. However the smooth decay of the X--ray light
curve is not observed in gamma--rays and the entire duration of the 40--700
keV burst is only $\sim 80$~s versus the $\sim 160$~s duration in
X--rays.

\begin{figure}[!t]
\epsscale{0.8}
\plotone{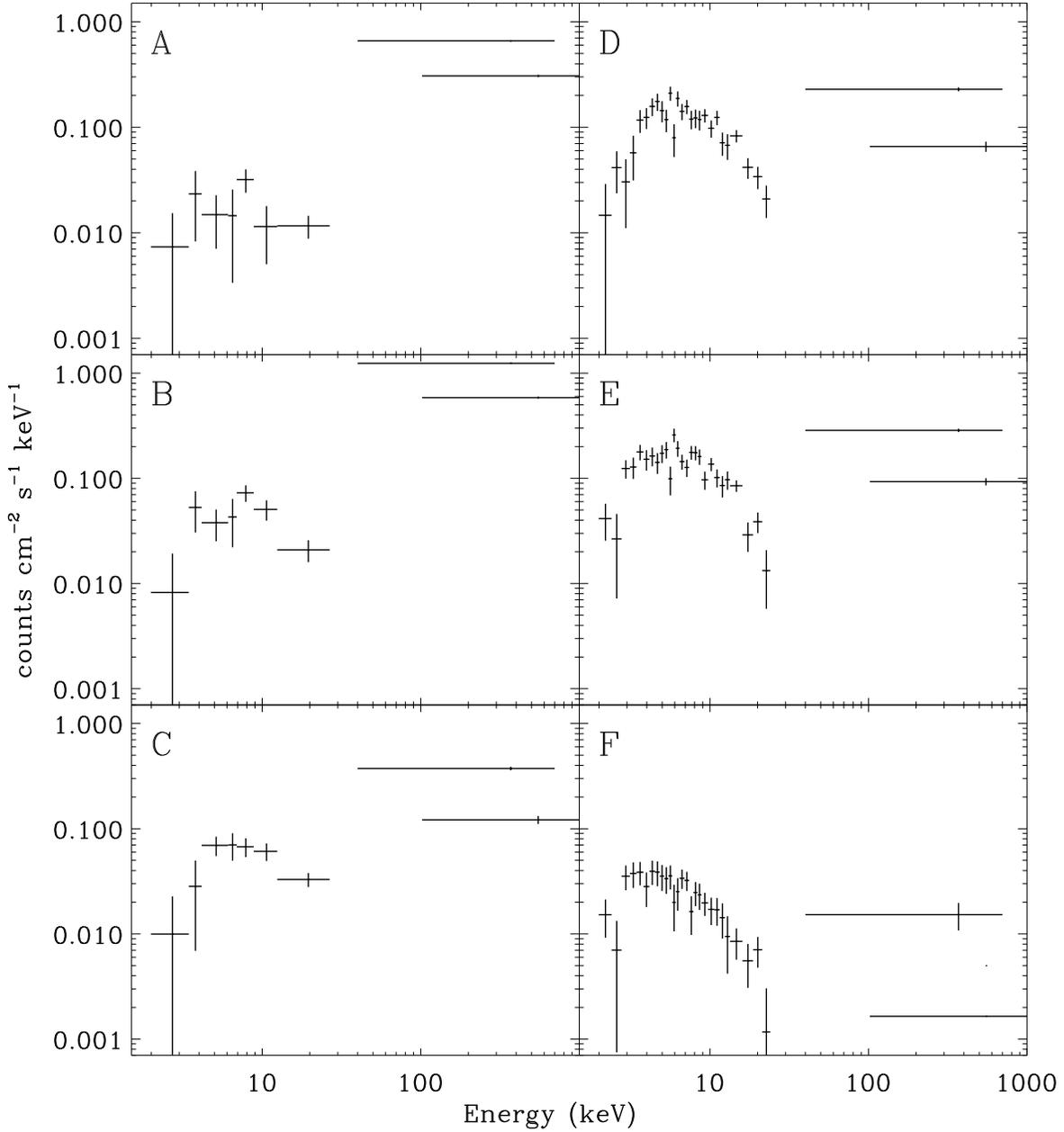}
\vspace{1.5cm}
\caption{Count spectra  of the burst in the time intervals A, B, C, D, E, and F.
For the spectral fits, we summed together the spectra A, B, and C, in order to
constrain the model parameters.}
\label{f:sp}
\end{figure}

For the study of the GRB spectral shape and its evolution with time, we 
extracted the WFC and GRBM data in the same time intervals and 
we simultaneously fitted the WFC and GRBM data. For GRBM
we used both types of data available: the 1~s counters and
the 128~s spectra.  Using the counters we had the possibility to
subdivide the GRB light curve in 6 adjacent time intervals A, B, C, D,
E, F of 18, 11, 10, 20, 20 and 90 s duration, respectively (see
Fig.~\ref{f:lc}) and to derive 2--700 keV count spectrum for each
of these intervals.  The obtained GRB count spectra are shown in
Fig.~\ref{f:sp}. Given that the first three spectra (A, B, and C) have
a low statistical quality in the 2-28 keV band, for the fitting
analysis they were summed together (interval ABC). Using 128~s GRBM 
transmitted spectra, we obtained two 2--700 keV spectra in the two time intervals
1 and 2, whose boundary time is marked in Fig.~\ref{f:lc}. 
The interval 1 covers the ABC interval above, while the interval 2 covers 
the intervals D, E and F.

Each of the four spectra (ABC, D, E, F) was fit with various
continuum models: a power-law ({\sc pl}), a power-law with an
exponential cutoff ({\sc cpl}), a broken power-law ({\sc bknpl}), a
smoothly broken power-law (or Band law, {\sc bl}, Band et al. 
1993\nocite{Band93}). All these models were assumed to be
photoelectrically absorbed by an equivalent--hydrogen column density
$N_{\rm H}$ to be determined (the {\sc wabs} model in {\sc xspec}).  
In the fits, the
normalisation factor of the GRBM data with respect to WFC was left
free to vary, checking that its best fit value was in the range (0.8
to 1.3) in which this parameter was found in the flight calibrations
with either celestial sources (e.g., Crab Nebula) or GRBs observed
with both the WFC-GRBM and the BATSE experiment aboard the {\it
Compton Gamma Ray Observatory} (e.g., Fishman et
al. 1994\nocite{Fishman94}).  The power--law model gave a completely
unacceptable fit of all count spectra either assuming the Galactic
column density along the GRB line of sight ($N_{\rm H}^{\rm G} = 6.1
\times 10^{20}$~cm$^{-2}$) or leaving the $N_{\rm H}$ free to vary in
the fit.  The fit with the Band law gave  in general unconstrained parameters.  
The other two models ({\sc cpl}), and {\sc bknpl}) resulted the most
suitable to describe the data. The fit results with these models
are reported in Table~\ref{t:fit-results}, along with the $\chi^2/dof$
value either in the case $N_{\rm H}$ is frozen to the Galactic
value ($N_{\rm H}^{\rm G}$) or in the case it is left free to vary in the fit. 

Apart from the
typical hard-to-soft evolution of the spectrum with time, the most
remarkable feature found is the value of the column density
$N_{\rm H}$, which results to be higher than the Galactic value 
in the first three time intervals ABC, D and E (see chance probabilities
obtained with F-test in Table~\ref{t:fit-results}) and it is consistent 
with 0, and thus with
$N_{\rm H}^{\rm G}$, in the interval F (2$\sigma$ upper limit reported
in Table~\ref{t:fit-results}). As can be seen from the F test
values, these results are independent of the spectral model chosen 
({\sc bknpl} or {\sc cpl}). We notice that, altough with 
lower significance, these results are confirmed with fits
using WFC data only.
Also the  analysis of the two GRBM 128~s spectra are consistent with 
these results. Both spectra were well fit with either a 
{\sc bknpl} or {\sc cpl}. In both
cases a column density $N_{\rm H} = N_{\rm H}^{\rm G}$ does not give
the best description of the data (see Table~\ref{t:fit-results}); a best
fit is obtained 
with  $N_{\rm H}$ values significantly higher than the Galactic column density.

Even if we cannot completely exclude a constantly high $N_{\rm H}$, 
the data are in favour of a decreasing column density.  
Indeed, summing together the data in the intervals 
ABC and D, we estimate an $N_{\rm H} = 1.0_{-0.3}^{+0.4}\times 10^{23}$~cm$^{-2}$, 
which is not consistent with the highest 2$\sigma$ upper limit of
the  $N_{\rm H}$ in the interval F
($4.4\times 10^{22}$~cm$^{-2}$) and it is only marginally consistent with 
its $3\sigma$ upper limit.  In addition, we find that the $N_{\rm H}$ 
centroid is continuously decreasing with time (see Fig.~\ref{f:NH}) and
the fit of the $N_{\rm H}$ values with a constant (weighted mean value $5.8\pm 1.1 
\times 10^{23}$~cm$^{-2}$) gives a $\chi^2/dof$ value  of 7.8/3, versus a value of 
0.65/2 in the case of an exponential law ($N_{\rm H}(t) = N_0 \exp (-t/t_0)$). 
Using the F-test \cite{Bevington69}, the probability  that a 
constant $N_{\rm H}$ value is the real parent distribution is  6\%.
Finally the comparison of the light curves in the 2--10 keV and 10--28 keV
energy bands (Fig.~\ref{f:NH}) are also suggestive of a variable $N_{\rm H}$. 
Assuming the above exponential law, the best fit parameters of the exponential law 
are $N_0 = (2.4 \pm 1.2) \times 10^{23}$ cm$^{-2}$ and $t_0 = 51 \pm 23$~s.

No evidence of an absorption edge associated with the variable $N_{\rm
H}$ is found in the spectra. The upper limit to the edge optical depth
$\tau$ depends on the energy assumed and on the time interval
considered. With the continuum parameters fixed to their best fit
values obtained with the {\sc cpl}, the $2\sigma$ $\tau$ upper limit
achieves the lowest value (0.2) at 5 keV in interval D and the highest
value (1.8) at 2 keV in the interval ABC. The other upper limits are
in between.

On the basis of the best fit results, peak fluxes and fluences
associated with the event were derived. The 2--10 keV and 40--700 keV
peak fluxes in 1~s time bin are $F_{2-10} = 2.1 \times 10^{-8}$~\ergcms and
$F_{40-700} = 1.4 \times 10^{-6}$~\ergcms, while the
fluences are given by $S_{2-10} = 9.1 \times 10^{-7}$~erg cm$^{-2}$,
and $S_{40-700} = 1.37 \times 10^{-5}$~erg cm$^{-2}$, with a softness
ratio $SR = S_{2-10}/S_{40-700} = 0.066$, which is a value typical of
a classic GRB (e.g., Frontera et al. 2000\nocite{Frontera00}).  Finally the 
total 2--700~keV fluence of the GRB is
given by $S_{2-700} = 1.9 \times 10^{-5}$~erg~cm$^{-2}$.

\subsection{Afterglow emission}
\label{s:aft}

Due to the weakness of the afterglow source (see Fig.~\ref{f:image}),
no statistically significant spectral information could be derived.
Assuming the average power-law photon index (1.95) derived from
the spectra of a large sample of GRB afterglows \cite{Frontera03a}, from
the MECS data we estimated the afterglow source flux in the two
observations of GRB000528, using an image extraction radius of 2$'$ and adopting
the MECS standard background level.  The resulting 2--10 keV mean flux
is $(1.6 \pm 0.4)\times 10^{-13}$~\ergcms ($1\sigma$ uncertainty)
during TOO1 and less than $5.2 \times 10^{-14}$~\ergcms ($2\sigma$
upper limit) for TOO2.  These fluxes can only give un upper limit to
the fading rate of the afterglow source. Assuming a power law for the
flux time behaviour ($F(t) \propto t^\delta$), with time origin at the
GRB onset, the $2\sigma$ upper limit to $\delta$ is $-0.74$, which is
not very constraining for the afterglow decay rate. Back extrapolated to
the time of the main event, this decay implies a flux much lower than
that measured with the WFC. A tighter constraint on $\delta$ can
be obtained with a reversed argument, imposing that -- as observed in
many \sax\ GRBs \cite{Frontera00,Frontera03a} -- the late part
of the prompt emission is correlated with the 
late afterglow emission, pointing to the former as early afterglow.  
In Fig.~\ref{f:aft} we show the results. Assuming  the GRB onset as origin of the 
afterglow onset time, we find that the smooth GRB decay starting at the 
beginning of the time interval F in Fig.~\ref{f:lc} is consistent with a 
power--law with index $\delta = -3.7\pm 0.3$.  As shown in Fig.~\ref{f:aft}
(left panel), 
its extrapolation to the time of the \sax\ NFI measurements gives a flux much 
lower than that measured. A different result is obtained if we fix as origin of 
the time coordinate for the afterglow the beginning of the time interval F. 
In this case, the 2--10 keV smooth decaying flux is  described 
by a broken power--law (a simple power-law gives is unacceptable,
$\chi^2/dof = 88/31$, see right-hand panel of Fig.~\ref{f:aft}), with
best fit parameters $\delta_1 = - 0.12 \pm 0.06$, $\delta_2 = - 1.33 \pm 0.13$ and 
break time $t_{break} = 16^{+1.9}_{-2.5}$~s. We notice that the second power--law 
index is that typical of X-ray afterglows \cite{Frontera03a} and that, when this
law is extrapolated to the epochs of the TOO1 and TOO2
measurements, it is found  consistent with the afterglow 
fluxes measured at these times. All that is  suggestive
of the fact that  the afterglow onset time is  delayed by $\sim 80$~s with 
respect to
the GRB onset time and that the final fading of the early afterglow of GRB000528 
occurs about 16~s after a slower fading.

\section{Discussion} 
\label{s:disc}
Two main results have been obtained from the present analysis of
the GRB000528 \sax\ data: the evidence of an early afterglow with onset time 
delayed with respect to the GRB onset, and the discovery of an absorption column 
density  significantly higher than the Galactic value during most of the burst 
prompt emission with evidence of a decreasing behavior with time. 
We discuss both.

\subsection{The early afterglow}

The finding that, assuming as time origin the beginning of interval F 
in Fig.~\ref{f:lc}, the extrapolation of fading law of the late prompt emission 
is consistent with the flux of the late afterglow, is highly suggestive 
of the expectations of the internal--external shocks model, during 
the fast cooling phase \cite{Sari95,Sari99}. 
%The late fading of the early afterglow tells us that the hydrodynamic
%evolution of the external shock  takes place in the thick shell regime, which is
%expected in the case of long GRBs, like GRB000528. 
The above derived early afterglow decay law and the photon spectrum of 
the last interval F (see Table~\ref{t:fit-results}) confirm this scenario.
Indeed, assuming a synchrotron emission, according to Sari \& Piran (1999),
the photon spectrum due to the fast cooling electrons is described by a set of
four power--laws, one $N(E)\propto \nu$ for $\nu<\nu<\nu_{sa}$, where $\nu_{sa}$
is the photon self-absorption frequency, the second $N(E)\propto \nu^{-2/3}$ for
$\nu_{sa}<\nu<\nu_c$, where $\nu_c$ is the cooling frequency, the third
$N(E)\propto \nu^{-3/2}$ for $\nu_c<\nu<\nu_m$, where $\nu_m$ is the characteristic
synchrotron frequency, and the last  $N(E)\propto \nu^{-p/2 - 1}$ for
$\nu>\nu_m$, where $p$ is the index of the electron  energy power-law distribution.
Comparing these expectations with our spectral results in the interval F (broken power-law
with low-energy photon index $\alpha = 1.27^{+0.36}_{-0.33}$, high-energy photon
index $\beta = 4.6_{-3.1}$, $E_{break} = 51^{+35}_{-29}$), we can see that $\alpha$
is consistent with the expected power law index (3/2) in the  $\nu_c<\nu<\nu_m$ frequency
band,
and thus that $\nu_m$ corresponds to  $E_{break}$. Unfortunately the high-energy
photon index $\beta$ is affected by a large uncertainty with the 90\% upper boundary
unconstrained, and thus it cannot be  efficiently used to constrain $p$. However
$p$ can be constrained by the measured light curve of the early afterglow.
%and it
%is found nicely similar to that expected at X--ray frequencies \cite{Sari99}. 
Indeed, in the case of the fast cooling phase \cite{Sari99}, at X--ray frequencies, 
the energy flux $F(t)$ is expected to increase as $t^2$ for $t<t_\gamma$, 
where $t_\gamma$ is the time in which the relativistic flow changes from a 
constant Lorentz factor into a decelarating phase, while $F(t)$ is expected to 
decay as $t^{-1/4}$ for $t_\gamma<t<t_m$, where $t_m$ is the time in which the 
typical synchrotron frequency cross the observed frequency; finally  $F(t)$ is 
expected to decay as $t^{1/2-3p/4}$ at later times ($t>t_m$). The light curve we 
observe is  nicely similar to that expected: it has first a temporal index 
$\delta_1 = - 0.12 \pm 0.06$, which is marginally consistent with that (-1/4) 
expected for $t<t_m$, then has a temporal index 
$\delta_2 = - 1.33 \pm 0.13$, which, compared with that expected ($1/2-3p/4$),
allows to estimate $p = 2.44 \pm 0.17$. This value of $p$ is consistent with
the estimated high-energy photon index of the {\sc bknpl}. 

Due to the thick shell regime, we are not capable to directly observe the onset time
of the afterglow and then to derive the initial Lorentz factor $\Gamma_0$ from the
decelaration time of the fireball $t_\gamma$, which depends on $\Gamma_0$ 
\cite{Sari99}. An attempt to evaluate $t_\gamma$ was done by fitting
the GRB light curve in the interval F with a broken power--law as above, 
in which this time $\delta_1$ is frozen to $-1/4$  and the other parameters,
$\delta_2$, $t_{break}$ and the onset time  $t_0$ of the afterglow with respect 
to the starting time of the interval F, were left free to vary. We found that 
the best fit is obtained with  $t_0 = -3^{+2}_{-5}$~s, $\delta_2 = 
- 1.29^{+0.11}_{-0.18}$, and $t_{break} = 17^{+3}_{-1}$~s. 
The fading law obtained with these parameters is still nicely consistent 
with the late afterglow measured flux. 
From this result, following Sari \& Piran (1999), we can get an estimate of the 
deceleration time $t_\gamma$ ($t_\gamma = -t_0$), and then an estimate of 
$\Gamma_0$ if we know the GRB redshift.  Placing 
GRB 000528 at $z=0.5$ (see below for a discussion), we obtain
$\Gamma_0 \sim 100 n_4^{-1/8}$  where $n_4$ is the density of the external
medium in units of $10^4$~cm$^{-3}$. We stress however that the derived value
of these parameters
should be taken as rough  estimates also because we adopted equations valid for an
unperturbed interstellar medium, which are not strictly applicable to
the thick shell case. In addition, it is worth considering that,
lacking a decay slope only constrained  from NFI observations, other possible
solutions could be considered. In spite of all this uncertainty we 
consider that outlined  a plausible scenario. 

\subsection{The column density behavior}

The other major result of the present spectral analysis of the GRB000528
prompt emission is the detection of an hydrogen-equivalent column
density significantly higher than the Galactic value. Even if,
as discussed, a constant value cannot be excluded, the likely  temporal
behavior of $N_{\rm H}$ is an exponential function with initial 
value $N_{\rm H}(0) \sim 2.4 \times 10^{23}$ cm$^{-2}$ (assuming a 
redshift $z= 0$) and decay constant $t_0 \sim 51$ s.
It is the second time, after the case of GRB980329
\cite{Frontera00,LazPer02}, that $N_{\rm H}$  such time behavior, 
although  evidences of
variable absorption during the prompt emission of GRBs have already been  
reported (GRB990705, Amati et al. 2000\nocite{Amati00},
GRB010222, in 't Zand et al. 2001\nocite{Zand01}, GRB010214, Guidorzi
et al. 2003\nocite{Guidorzi03}).  As discussed in
Section~\ref{s:intro}, in the case of GRB980329, the $N_{\rm H}$
behavior was found consistent with a photoionization
process, which is expected when a GRB occurs within a cloud of
initially cold gas. The cloud mass density
was found to be $n\sim 4.5 \times 10^5$~cm$^{-3}$ for a composition typical of
interstellar matter, and the radial distribution was consistent either
with a uniform sphere of radius $R_{sphere} = 0.13$~pc, or a shell at
distance $R_{shell} = 0.066$~pc and width $\Delta R=0.1 R_{shell}$. A
marginal but not conclusive preference was given to the shell geometry
\cite{LazPer02}. 

\begin{figure}[!t]
\epsscale{0.8}
\plotone{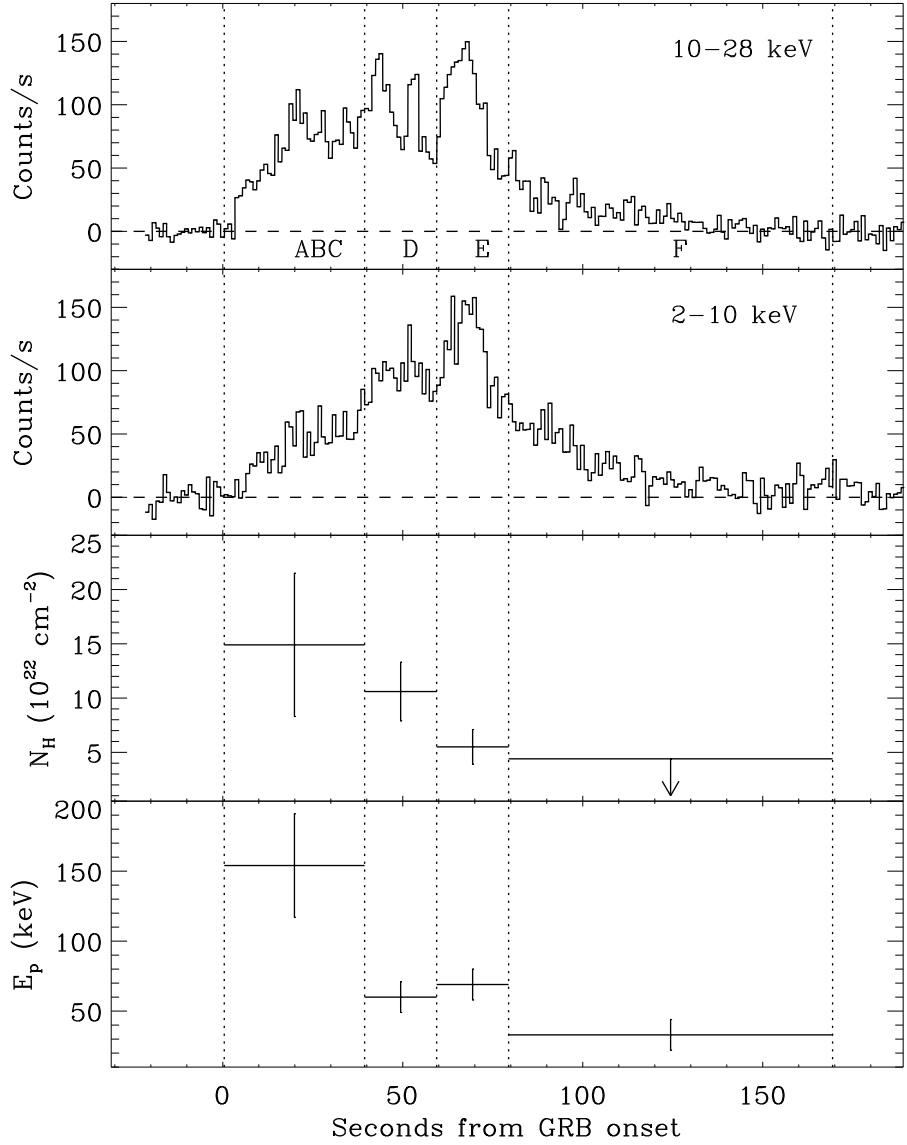}
\vspace{1.5cm}
\caption{Time behaviour of the hydrogen equivalent column density and
peak energy of the $EF(E)$ spectrum. Also the 2--10 keV and 10--28 keV 
light curves are shown.}
\label{f:NH}
\end{figure}

For GRB000528 we tested the same model with two important improvements
in the simulation of the photoionization process.  First, we have
taken into account the true gamma-ray light curve of the GRB, while in
the case of GRB980329 a constant flux was assumed to be incident on
the cloud.  Second, we have taken into account the cosmological effect
on both the time dilation of the light curves and  the measured
column density: the fit was performed for three different values of
$z$ (0.1, 0.3, 0.5). Consequently, the $N_{\rm H}$ values shown in
Fig.~\ref{f:NH} were corrected to take into account the redshift
effect on the column density according to the relation $N_{\rm H}(z) =
N_{\rm H}(0)\,(1+z)^{2.5}$, which is valid in the redshift range $0 <
z < 4$ \cite{LazPer02}.  The results of the model fit to the data are
shown in Fig.~\ref{f:lazzati} and in Table~\ref{t:lazzati}, for the
three values of $z$. From the $\chi^2$ values shown in
Table~\ref{t:lazzati}, also in this case the two cloud geometries
(sphere or shell) cannot be significantly disentangled, even if it
appears, also in this case, a slight preference for the shell
geometry.  Assuming a spherical geometry, the radius of the 
cloud increases with $z$, from $8.9 \times 10^{16}$~cm to $5.0 \times
10^{17}$~cm, while in the case of a shell, the shell distance increases
with $z$ from $5.6 \times 10^{16}$~cm to $1.8 \times 10^{17}$~cm (see
Table~\ref{t:lazzati}). 
Also the initial column density is sensitive to the redshift
assumed, ranging from from $6.3 \times 10^{23}$~cm$^{-2}$ to $1.0
\times 10^{24}$~cm$^{-2}$ in the first case, and from $3.5 \times
10^{23}$~cm$^{-2}$ to $8.9 \times 10^{23}$~cm$^{-2}$ in the second
case\footnote{An initial column density larger than
$10^{24}$~cm$^{-2}$ has not been considered since it would affect the
$\gamma$-ray lightcurve of the GRB.}. However, independently of the
redshift and cloud geometry, the best fit values are all consistent
with overdense regions of molecular clouds where star formation takes
place, as also found for GRB980329 \cite{LazPer02}. 

The amount of
mass required to explain the variable absorption ranges between few to
few hundreds solar masses. This mass range is consistent with the
masses ejected by Wolf-Rayet stars, the candidate progenitors of GRB
explosions. If the mass were distributed as a wind, with an $r^{-2}$
profile, most of the absorption would be at small radii, and the
column density would be expected to vanish quickly. Lazzati \& Perna
(2002) simulated the evolution of absorption from a stellar wind,
finding that even a Thompson thick wind would vanish in less than a
second (see their Fig.~6). This does not mean that our observations
provides evidence against a stellar wind. It means that this kind of
observation is not sensitive to the presence of a wind. Nevertheless,
the observation can be interpreted as a circumstantial evidence for
the stellar origin of the progenitor. As recently discussed by
Chevalier et al. (2004), the interaction of the wind with the
surrounding ISM creates a dense and cold shell of material consistent
with the shell inferred from our results.

The above results suggest that the high density of the circumburst
medium may be responsible for the unsuccessful search of an optical
counterpart (Jensen et al. 2000, Palazzi et al. 
2000\nocite{Jensen00,Palazzi00}). In most of the best fit
simulations (see Fig.~\ref{f:lazzati}), a sizable column density of
the order of several$\times10^{21}$~cm$^{-2}$ remains after the GRB
prompt phase has finished. Dust associated to this residual column
density may well be responsible for the optical obscuration, even for
a Galactic dust to gas ratio. In addition, Perna \& Lazzati
(2002)\nocite{perlaz02} showed that depending on the cloud geometry
and GRB spectrum, the dust to gas ratio at the end of the burst phase
may be even larger than the initial one. 

\begin{figure}[!t]
%\epsscale{0.8}
\plotone{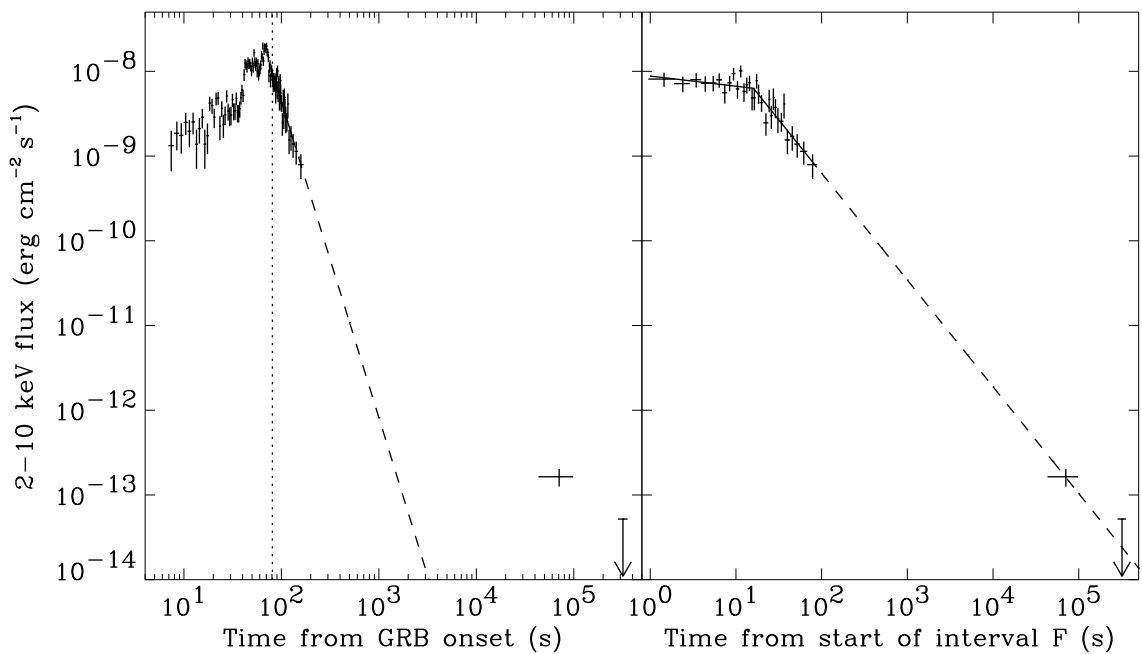}
\vspace{0.5cm}
\caption{2-10 keV fading curve of GRB000528. 
The afterglow upper limit is at $2\sigma$ level. The vertical dotted line in
the lfet panel shows the beginning of the interval F. 
{\it Left panel:} light curve of the GRB and its afterglow assuming as origin the
onset time of the GRB. The best fit of the light curve  in the interval F
and its extrapolation at later times is also shown. 
{\it Right panel:} The light curve in the interval F and its extrapolation 
at later epochs, when the beginning of the interval F is assumed as origin of
the afterglow onset time.}
\label{f:aft}
\end{figure}

\begin{figure*}[!t]
\epsscale{0.8}
%\plotone{lazzati_model.ps}
\centerline{\psfig{figure=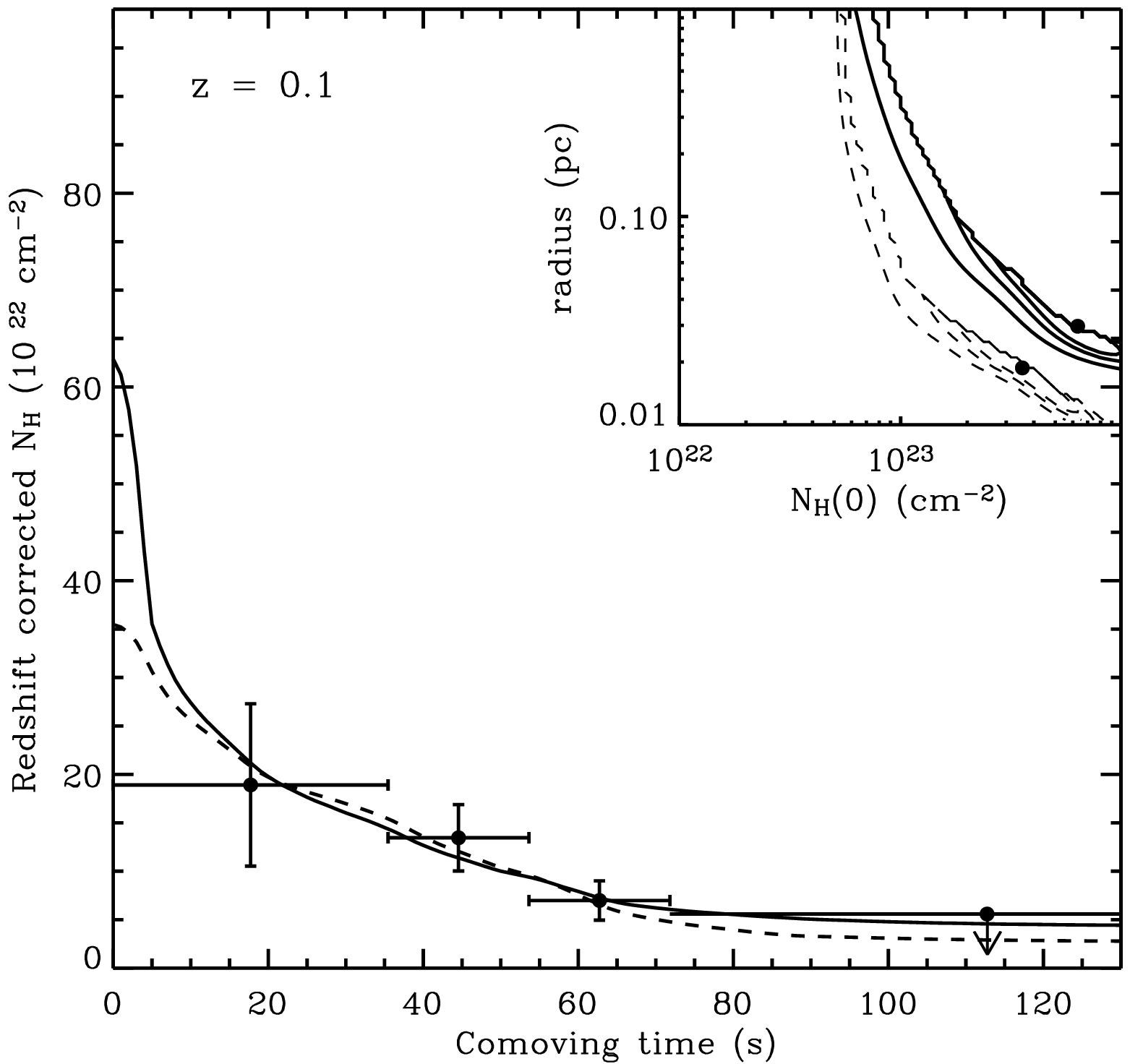,width=5.5cm,angle=0}\psfig{figure=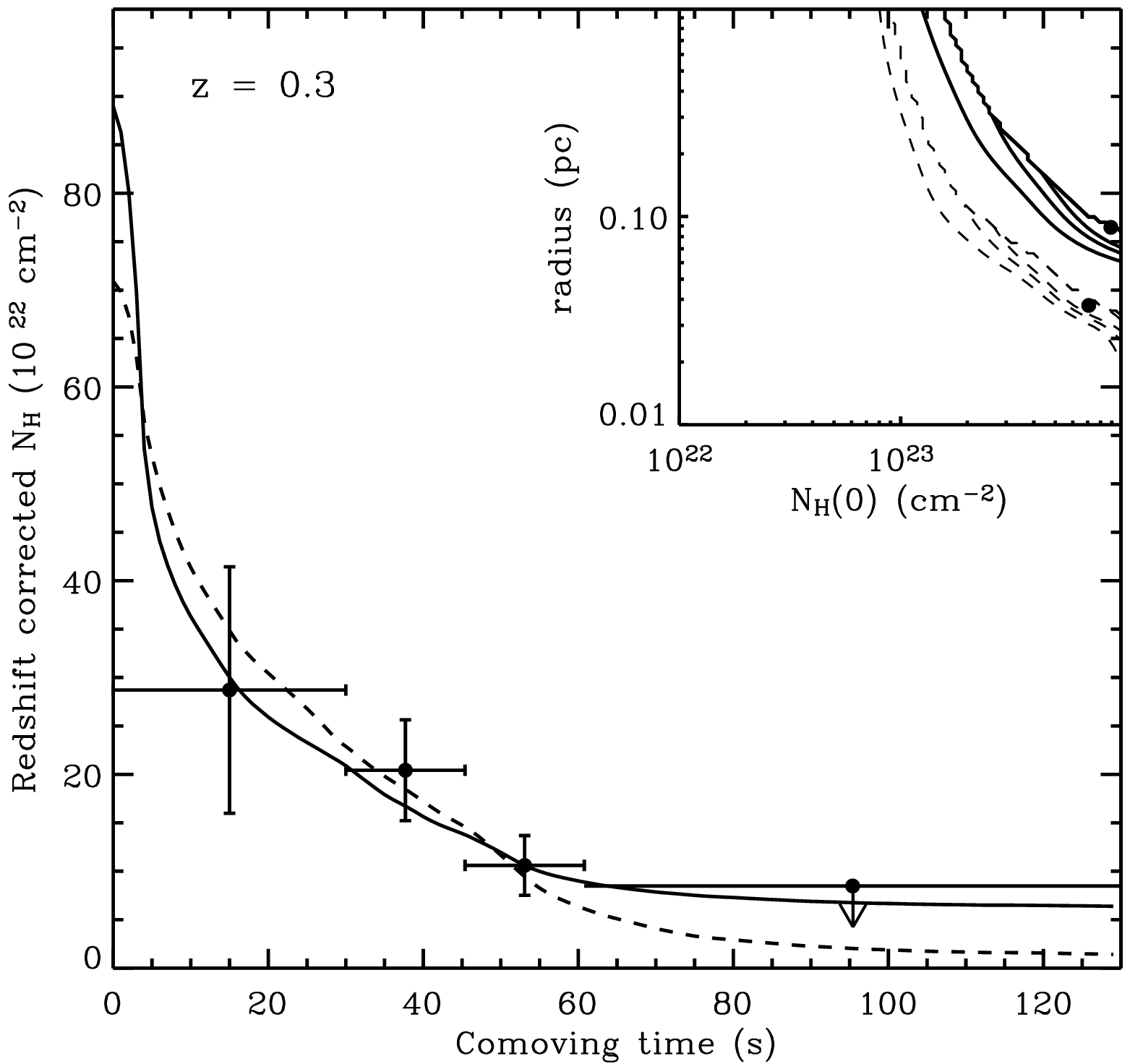,width=5.5cm,angle=0}\psfig{figure=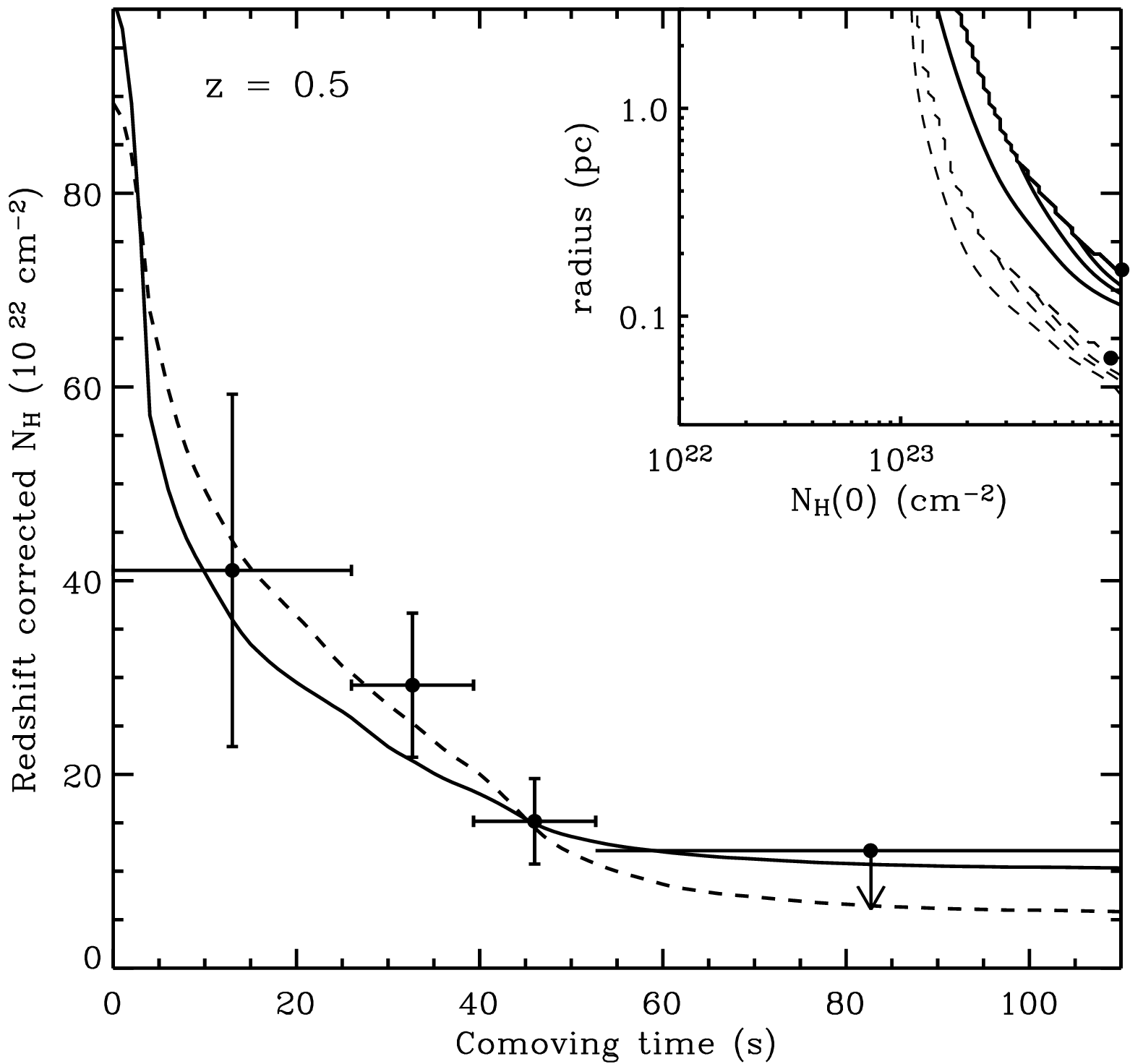,width=5.5cm,angle=0}}
\vspace{0.5cm}
\caption{Redshift corrected column density as a function of the time in the
rest frame of the burst for three different redshift assumed. Also the
best fit model (see text) for the two cloud geometries: sphere ({\it continuous 
line}), and shell ({\it dashed line}).
In the inset the corresponding best fit values (initial redshift corrected 
$N_{\rm H}$ and cloud radius) are shown for two geometries along with 
the $1\sigma$, 90\% and 99\% confidence contours. The 90\% and the 99\% 
contours are not closed.}
\label{f:lazzati}
\end{figure*}

Remarkably, the slope shown by the X--ray afterglow emission is that 
typical of a GRB with constant density \cite{Sari98},
which is the assumption for the cloud in the Lazzati \& Perna (2002) model.   
We do not find evidence of a temporal break at least until 1 day from the 
main event.  Also the afterglow upper limit in TOO2 is consistent 
with the same temporal slope (see Fig.~\ref{f:aft}). If, in spite of 
that, we assume that a break in the light curve occurs 2 days after the burst, 
in the scenario of the internal--external shock model, the break could be a
consequence of a transition to a non relativistic phase of the shock
\cite{Dai99}, due to a high density medium. Using the Dai \& Lu (1999) 
model, we can get an estimate of this density.  Assuming a Lorentz
factor $\gamma = 1$ at $t = 2$~days, from Dai \& Lu (1999), we find
that $E_{54} n_5^{-1/8}[(1+z)/2.6]^{3/8} = 0.65$, where $E_{54}$ is
the total isotropic-equivalent energy $E_{rad}$ in units of
$10^{54}$~erg, $n_5$ is the medium density $n$ in units of $10^5$~
cm$^{-3}$ and $z$ is the GRB redshift.  Assuming $z = 0.5$, which is
also consistent with the minimum value inferred from the Amati et
al. (2002)\nocite{Amati02} relation ($E_p$--$E_{rad}$) for a 20\%
dispersion of the data points around the best fit, for a standard
cosmology with $H_0 = 71 $~km~s$^{-1}$~Mpc$^{-1}$, $\Omega_m = 0.3$ and
$\Omega_\Lambda = 0.7$ \cite{Spergel03}, we find $E_{54} = 0.014$ and then a medium
density of $n = 4.5 \times 10^4$~cm$^{-3}$. This value is somewhat
smaller than the best fit densities derived from the fit of the column
density evolution. However, as shown in the insets of
Fig.~\ref{f:lazzati}, the data do not allow to constrain the cloud
geometry with high accuracy, and lower density solutions are
consistent with the $N_{\rm H}$ evolution. In addition, given that 
late afterglow prevents us to infer a fading law independently  of the
late prompt emission time behavior, we cannot exclude 
consistent afterglow solutions with a break before the first NFI
observation, thus relaxing the upper limit on the ISM density. At any
event, for the best fit $N_{\rm H}$, the cloud geometry and density
are consistent with the characteristics of massive cloud cores (Plume
et al. 1997), and therefore with the progenitor being a massive,
short-lived star.

Continuous, early time monitoring of the time variable opacity 
(both in single lines and in continuum) has the potential
to constrain the run of density with radius (Lazzati et al. 2001b),
and hence to trace the mass loss history of the progenitor star.
These observations are expected to be partly covered by the upcoming satellite 
{\em Swift}, with its repointing capabilities of the X-ray and optical
detectors in several tens of seconds. A mission which would be capable to fully
monitor the GRB opacity since the burst onset is the proposed 
{\em Lobster--ISS} for the International Space Station, a
sensitive all-sky monitor with focusing optics from 0.1 to 3.5 keV 
\cite{Fraser02} paired with a GRB monitor for higher photon energies (from 2 keV 
to several hundreds of keV) \cite{Frontera03b,Amati04}. 
When measurements of    
variable absorption in X-rays are complemented with those in the optical
band, one can further put powerful constraints on the properties of dust 
in the environment of the burst (Perna, Lazzati \& Fiore 2003)\nocite{PLF03}.

\acknowledgements

We wish to acknowledge  Jean in 't Zand for his 
help in the WFC data reduction. This research was supported by 
both the Italian Space Agency (ASI) and 
Ministry of the University and Research of Italy (COFIN funds 2003). \sax\ was
a joint program of the Italian Space Agency and the Netherlands Agency for Aerospace
Programs.

\clearpage

\begin{deluxetable}{ccccccccc}
\small
%\footnotesize
%\scriptsize
\tablewidth{0pt}
\tablenum{1}
\tablecaption{Spectral parameters of the {\sc bknpl} and {\sc cpl} best fit models}
\tablehead{
Interval  & Model & $N_{\rm H}$ & $\alpha$   & $\beta$  & $E_{b}$  & 
$E_c$  &  $\chi^2/dof$ & F test \tablenotemark{(a)} \\
	  & 	& ($10^{22}$~cm$^{-2}$) &    &          & (keV)	  & (keV)  &		
&               
} 
\startdata
ABC & {\sc bknpl} &  [0.061]	  & $0.02_{-0.24}^{+0.20}$ & $2.10 _{-0.20}^{+0.10}$ & $46_{-9}^{+12}$ &
       &         11.2/5 &	  \\
    & {\sc bknpl} & $13.1_{-8.3}^{+18.8}$ & $0.54_{-0.46}^{+0.42}$ & $2.13_{-0.22}$ & $56_{-18}^{+99}$ &
       &         5.6/4 & 0.12	 \\
     & {\sc cpl}   & $14.9_{-8.3}^{+12.6}$ & $0.36_{-0.23}^{+0.24}$ &    &      & 
$80_{-18}^{+22}$    &    5.6/5 & 0.033 \\
D   & {\sc bknpl} & [0.061]       & $0.26_{-0.20}^{+0.18}$ & $2.28_{-0.15}^{+0.26}$ & $15_{-3}^{+7}$ &
                & 53/26 &	\\
    & {\sc bknpl} & $10.6_{-4.1}^{+5.8}$ & $1.05_{-0.24}^{+0.27}$ & $2.88_{-0.37}^{+0.39}$ & $36_{-11}^{+12}$ &
                & 28.7/25 &  $1.2 \times 10^{-4}$	\\
    & {\sc cpl}   & $10.6_{-4.0}^{+5.0}$ & $0.91_{-0.25}^{+0.25}$ & 		      &		  &
$55_{-14}^{+23}$     & 28.8/26  & $2.6 \times 10^{-5}$	\\
E   & {\sc bknpl} & [0.061]       & $0.64_{-0.10}^{+0.11}$ & $2.66_{-0.30}^{+0.28}$ & $33_{-7}^{+7}$  &
                & 54.8/26 &	\\
    & {\sc bknpl} & $5.3_{-2.5}^{+3.2}$ & $1.01_{-0.20}^{+0.21}$ & $2.64_{-0.28}^{+0.30}$ & $39_{-10}^{+12}$ &
                & 40/25 &  $5.2 \times 10^{-3}$  \\
    & {\sc cpl}   & $5.5_{-2.4}^{+3.1}$ & $0.93_{-0.19}^{+0.19}$ & 		      &		  &
$67_{-17}^{+21}$     & 40.6/26  & $2.4 \times 10^{-3}$	\\
F   &  {\sc bknpl} & [0.061]       & $1.02_{-0.18}^{+0.18}$ & $4.2_{-2.8}$ 
\tablenotemark{(b)} & $38_{-29}^{+23}$  &
                & 15.6/26 &	\\
    & {\sc bknpl} & $<4.2$ ($2\sigma$) & $1.27_{-0.33}^{+0.36}$ & $4.6_{-3.1}$
\tablenotemark{(b)} & $51_{-29}^{+35}$ &
                & 13/25 & 0.24	 \\
    & {\sc cpl}   & $<4.4$ ($2\sigma$) & $0.95_{-0.45}^{+0.50}$ & 		      &		  &
$31_{-12}^{+29}$     & 13.5/26  &0.20\\	
\\
 1   & {\sc bknpl} & [0.061]       & $0.37_{-0.10}^{+0.15}$ & $2.26_{-0.17}^{+0.22}$ & $104_{-13}^{+14}$  &
                & 119/76 &	\\
    & {\sc bknpl} & $26.7_{-15.6}^{+23.7}$ & $0.84_{-0.15}^{+0.14}$ & $2.49_{-0.24}^{+0.30}$ & $130_{-18}^{+17}$ &
                & 105/75 &  $2.1 \times 10^{-3}$  \\
    & {\sc cpl}   & $10.1_{-8.0}^{+14.9}$ & $0.29_{-0.14}^{+0.15}$ & 		      &		  &
$103_{-13}^{+16}$     & 98.9/76  & $8.0 \times 10^{-2}$	\\
 2   & {\sc bknpl} & [0.061]       & $0.58_{-0.11}^{+0.09}$ & $2.24_{-0.13}^{+0.13}$ & $18_{-4}^{+3}$  &
                & 80.9/66 &	\\
    & {\sc bknpl} & $5.3_{-1.9}^{+2.2}$ & $1.00_{-0.14}^{+0.14}$ & $2.33_{-0.13}^{+0.12}$ & $33_{-6}^{+6}$ &
                & 60.7/65 &  $1.6 \times 10^{-5}$  \\
    & {\sc cpl}   & $7.0_{-1.8}^{+2.2}$ & $1.08_{-0.10}^{+0.11}$ & 		      &		  &
$81_{-13}^{+18}$     & 56.2/66  & $5.5 \times 10^{-12}$	\\
\enddata
\tablenotetext{a}{The column reports the chance probability values obtained with
the F-test}
\tablenotetext{b}{ The 90\% upper boundary value is unconstrained by the fit}
\label{t:fit-results}
\end{deluxetable}

\begin{deluxetable}{ccccc}
%\small
%\footnotesize
%\scriptsize
\tablewidth{0pt}
\tablenum{2}
\tablecaption{Best fit parameters of the Lazzati \& Perna (2002) model for
evolution of the column density. Error on the parameters are not quoted 
given the complex covariances (see Fig.~\ref{f:lazzati}).}
\tablehead{
Cloud Geometry  & Redshift & $N_{\rm H}(0)$ & $Radius$ & $\chi^2/dof$ \\
	        &         & ($10^{23}$~cm$^{-2}$) & (pc)   &          \\ 
} 
\startdata
sphere       & 0.1         &  6.31          &  0.029 	  &  0.65     \\
             & 0.3	   &  8.91          &  0.086 	  &  0.55     \\
	     & 0.5	   &  10.0	    &  0.162	  &  0.97     \\	
shell	     & 0.1	   &  3.55	    &  0.018	  &  0.15     \\	
	     & 0.3	   &  7.08	    &  0.036	  &  0.35     \\	
	     & 0.5	   &  8.91	    &  0.061	  &  0.28   
\enddata
\label{t:lazzati}
\end{deluxetable}

\clearpage

% Figure 1

% Figure 2

%
% Figure 3
%

%
% Figure 4
%

%
% Figure 5
%

\clearpage

%
%Figure 6
%

\begin{thebibliography}{99}

\bibitem[Amati et al. 2000]{Amati00}
Amati, L. \etal 2000, Science, 290, 953
%
\bibitem[Amati et al. 2002]{Amati02}
Amati, L. \etal 2002, \aap, 390, 81 
%
\bibitem[Amati et al.\ 2004]{Amati04}
Amati, L. \etal 2004, paper to be presented at the
COSPAR Symposium, Paris, July 2004.
%
\bibitem[Arnaud 1996]{Arnaud96}
Arnaud, K. A. 1996, in ASP Conf. Series 101, Astronomical Data
Analysis Software and Systems V, ed.\ G. H. Jacoby \& J. Barnes
(San Francisco: ASP), 17
%
\bibitem[Band et al.\ 1993]{Band93}
Band, D. \etal 1993, \apj, 413, 281 
%
\bibitem[Berezhiani et al.\ 2003]{Berezhiani03}
Berezhiani, Z., Bombaci, I., Drago, A., Frontera, F., \& Lavagno, A.,
2003, \apj, 586, 1250
%
\bibitem[Berger \& Frail 2000]{Berger00}
Berger, E. \& Frail, D. A. 2000, GCN 686
%
\bibitem[Bevington 1969]{Bevington69}
Bevington, P.R. 1969, {\it Data Reduction and Error Analysis for the
Physical Sciences} (McGraw-Hill, New York), p. 195.
%
\bibitem[B\"ottcher et al.\ 1999]{Boettcher99}
B\"ottcher, M., Dermer, D.D., Crider, W., Liang, E.P. 1999, \aap, 343, 111
%
\bibitem[Chevalier et al. 2004]{Chev04}
Chevalier, R. A., Li, Z., Fransson, C., 2004, ApJ, 606, 369
%
\bibitem[Connors \& Hueter 1998]{Connors98}
Connors, A. \& Hueter, G. J. 1998, \apj, 501, 307
%
\bibitem[Dai \& Lu 1999]{Dai99}
Dai, Z.G. \& Lu, T. 1999, \apj, 519, L155
%
%\bibitem[Dickey \& Lockman 1990]{Dickey90}
%Dickey, J.M., \& Lockman, F.J. 1990, ARA\&A, 28, 215
%
\bibitem[Fishman et al.\ 1994]{Fishman94}
Fishman, G.J. \etal 1994, \apjs, 92, 229
%
\bibitem[Fraser et al. 2002]{Fraser02}
Fraser, G. \etal 2002, Spie Proc. 4497, 115
%
\bibitem[Frontera et al.\ 1997]{Frontera97}
Frontera, F. \etal  1997, \aaps, 122, 357
%
\bibitem[Frontera et~al.\ 2000]{Frontera00}
Frontera, F. \etal 2000, \apjs, 127, 59 
%
\bibitem[Frontera 2003a]{Frontera03a}
Frontera, F. 2003a, in {\it Supernovae and gamma Ray Bursters}, 
ed. K. W. Weiler, (Springer, Berlin Heidelberg), p. 317
%
\bibitem[Frontera et~al.\ 2003b]{Frontera03b}
Frontera, F. \etal 2003b, talk presented at the National 
Conference on Compact Objects, Monteporzio, December 2003 
{\it (http://argos.mporzio.astro.it/cnoc3/cnoc\_files/schedula.html)}
%
\bibitem[Frontera et~al.\ 2004a]{Frontera04a}
Frontera, F. 2004a, in Proc.s of the 3rd Workshop on {\it Gamma Ray Bursts  in the
Afterglow Era}, edited by M. Feroci, F. Frontera, N. Masetti \& L. Piro
(ASP Publ.), vol. 312, in press.
%
\bibitem[Frontera et~al.\ 2004b]{Frontera04b}
Frontera, F. \etal 2004, \apj, submitted
%
\bibitem[Frontera et~al.\ 2004c]{Frontera04c}
Frontera, F. \etal 2004c, in preparation 
%
\bibitem[Gandolfi 2000]{Gandolfi00}
Gandolfi, G. 2000, BeppoSAX Mails N. 00/10 and 00/11
%
\bibitem[Guidorzi et al.\ 2000]{Guidorzi00}
Guidorzi, C., Montanari, E., Frontera, F. \& Feroci, M. 2000, GCN 675
%
\bibitem[Guidorzi et al.\ 2003]{Guidorzi03}
Guidorzi, C. \etal 2003, \aap, 401, 491
%
\bibitem[Hijorth et al.\ 2003]{Hijorth03}
Hijorth, J. \etal 2003, Nature, 423, 847
%
\bibitem[Hurley et al.\ 2000]{Hurley00}
Hurley, K. \etal 2000, GCNs 681, 700
%
\bibitem[Jager et~al.\ 1997]{Jager97}
Jager, R., \etal  1997, \aaps, 125, 557
%
\bibitem[Jensen et al.\ 2000]{Jensen00}
Jensen, B.L. \etal 2000, GCN 674
%
\bibitem[Kuulkers et al.\ 2000]{Kuulkers00}
Kuulkers, E. \etal 2000, GCN 700
%
\bibitem[Lazzati et al. 2001a]{laz2001a}
Lazzati, D., Ghisellini, G., Amati, L., Frontera, F., Vietri, M. \&
Stella, L., 2001a, ApJ, 556, 471
%
\bibitem[Lazzati et al. 2001b]{laz2001b}
Lazzati, D., Perna, R. \& Ghisellini, G. 2001b, MNRAS, 325, L19
%
\bibitem[Lazzati \& Perna 2002]{LazPer02}
Lazzati, D. \& Perna, R., 2002, \mnras, 330, 383
%
\bibitem[Lazzati et al. 2002]{laz2002}
Lazzati, D., Ramirez-Ruiz, E. \& Rees, M. J., 2002, ApJ, 572, L7
%
\bibitem[Palazzi et al.\ 2000]{Palazzi00}
Palazzi, E. \etal 2000, GCN 691
% 
%\bibitem[Morrison \& McCammon 1983]{Morrison83}
%Morrison R. and McCammon, D. 1983, \apj, 270, 119
%
\bibitem[Paczynski 1998]{Paczynski98} 
Paczynski, B. 1998, \apj, 494, L45
%
\bibitem[Perna \& Loeb 1998]{perlo98}
Perna R. \& Loeb A., 1998, \apj, 501, 467
%
\bibitem[Perna \& Lazzati 2002]{perlaz02}
Perna, R. \& Lazzati, D., 2002, \apj, 580, 261
%
\bibitem[Perna, Lazzati \& Fiore 2003]{PLF03}
Perna, R., Lazzati, D. \& Fiore, F. 2003, \apj, 585, 775
%
\bibitem[Plume et al. 1997]{plume97}
Plume R., Jaffe D. T., Evans N. J., Mart{\'\i}n-Pintado J.,
	G\'omez-Gonz\'alez J., 1997, \apj, 476, 730
%
\bibitem[Piro et al.\ 2000]{Piro00}
Piro, L. \etal 2000, Science, 290, 955
%
\bibitem[Reeves et al.\ 2002]{Reeves02}
Reeves, J.N. \etal 2002a, \nat, 416, 512
%
\bibitem[Reeves et al.\ 2003]{Reeves03}
Reeves, J.N. \etal  2003, \aap, 403, 463
%
%\bibitem[Sari 1997]{Sari97}
%Sari, R. 1997, \apj, 489, L37
%
\bibitem[Sari \& Piran 1995]{Sari95} 
Sari, R. \& Piran, T. 1995, ApJ, 455, L143
%
\bibitem[Sari et al.\ 1998]{Sari98}
Sari, R., Piran, T., \& Narayan, R. 1998, \apj, 497, L17
%
\bibitem[Sari \& Piran 1999]{Sari99} 
Sari, R. \& Piran, T. 1999, ApJ, 520, 641
%
\bibitem[Spergel et al.\ 2003]{Spergel03}
Spergel, D.N. \etal 2003, \apjs, 148, 161
%
\bibitem[Stanek et~al.\ 2003]{Stanek03}
Stanek, K. Z. \etal 2003, \apj, 591, L17
%
\bibitem[Vietri \& Stella 1998]{Vietri98}
Vietri, M. \&  Stella, L. 1998, \apj, 507, L45
%
%\bibitem[Woosley 1993]{W93}
%Woosley, S. 1993, ApJ, 405, 273
%
\bibitem[in 't Zand et al.\ 2000]{Zand00}
in 't Zand \etal 2000, GCN 677
%
\bibitem[in't Zand et al. 2001]{Zand01}
in't Zand, J.J.M., Kuiper, L. et al. 2001, \apj, 559, 710 
%
\end{thebibliography}
\end{document}